\documentclass[%
reprint,
superscriptaddress,
preprintnumbers,
nofootinbib,
amsmath,amssymb,
aps,
prd,
]{revtex4-1}

\usepackage{hyperref}
\hypersetup{colorlinks=true, citecolor=blue, urlcolor=blue, linkcolor=blue}
\usepackage[T1]{fontenc}
\usepackage{float}      		
\usepackage{dcolumn}    		
\usepackage{bm}         		
\usepackage{graphicx}		
\usepackage{amsmath}		
\usepackage{amssymb}		
\usepackage{aas_macros}
\usepackage{xcolor}
\usepackage{newtxtext,newtxmath}
\usepackage[linesnumbered,ruled,vlined]{algorithm2e}
\usepackage{algpseudocode}
\usepackage{tikz}
\usetikzlibrary{fit,calc}
\usepackage{makecell}

\begin{document}
\title{Fast Generation of Weak Lensing Maps with Analytical Point Transformation Functions}

\author{Kunhao Zhong}\email{kunhaoz@sas.upenn.edu}
\affiliation{Department of Physics and Astronomy,
University of Pennsylvania, Philadelphia, PA 19104, USA}

\author{Gary Bernstein}
\affiliation{Department of Physics and Astronomy,
University of Pennsylvania, Philadelphia, PA 19104, USA}

\author{Supranta S. Boruah}
\affiliation{Department of Physics and Astronomy,
University of Pennsylvania, Philadelphia, PA 19104, USA}

\author{Bhuvnesh Jain}
\affiliation{Department of Physics and Astronomy,
University of Pennsylvania, Philadelphia, PA 19104, USA}

\author{Sanjit Kobla}
\affiliation{Department of Physics and Astronomy,
University of Pennsylvania, Philadelphia, PA 19104, USA}

\date{\today}

\begin{abstract}
Nonlinear cosmological fields like galaxy density and lensing convergence can be approximately related to Gaussian fields via analytic point transforms. The lognormal transform (LN) has been widely used and is a simple example of a function that relates nonlinear fields to Gaussian fields. We consider more accurate General Point-Transformed Gaussian (GPTG) functions for such a mapping and apply them to convergence maps. We show that we can create maps that preserve the LN's ability to exactly match any desired power spectrum but go beyond LN by significantly improving the accuracy of the probability distribution function (PDF). With the aid of symbolic regression, we find a remarkably accurate GPTG function for convergence maps: its higher-order moments, scattering wavelet transform, Minkowski functionals, and peak counts match those of N-body simulations to the statistical uncertainty expected from tomographic lensing maps of the Rubin LSST 10 years survey. Our five-parameter function performs 2 to 5$\times$ better than the lognormal. We restrict our study to scales above about 7 arcmin; baryonic feedback alters the mass distribution on smaller scales. We demonstrate that the GPTG can robustly emulate variations in cosmological parameters due to the simplicity of the analytic transform. This opens up several possible applications, such as field-level inference, rapid covariance estimation, and other uses based on the generation of arbitrarily many maps with laptop-level computation capability. 

\end{abstract}

\maketitle

\section{Introduction}

Gravitational lensing, a phenomenon predicted by general relativity, occurs when the trajectory of photons is distorted as they propagate through the inhomogeneous universe with varying gravitational potential. Weak gravitational lensing refers to the subtle distortions of source images due to the mass distribution of lenses. By coherently extracting these distortions across a large number of sources, we can map the large-scale structure of the universe independently of assumptions about how observable baryonic matter traces the underlying dark matter field. These maps, often called weak lensing mass maps in the literature, provide a valuable tool for probing the growth and geometry of the universe through statistical analysis. For a comprehensive review, see Ref.~\cite{Mandelbaum:2017jpr}.

Current Stage-III imaging surveys have yielded promising results for understanding the evolution and fundamental physics of the universe. These surveys include the Dark Energy Survey (DES)~\cite{DES:2016jjg,DES:2017qwj,DES:2021bvc}, the Kilo-Degree Survey (KiDS)~\cite{Kuijken:2015vca,Heymans:2020gsg,KiDS:2020suj}, the Hyper Suprime-Cam Subaru Strategic Program (HSC)~\cite{2023arXiv230400701D, 2023arXiv230400702L, 2023arXiv230400705S}. The upcoming Stage-IV surveys will provide weak lensing data of unprecedented accuracy, while simultaneously presenting challenges for drawing unbiased conclusions. The two-point correlation function has been the most extensively studied statistical tool, with over a decade of research focused on systematics and synthetic checks. However, this method only extracts Gaussian information from weak lensing maps, potentially overlooking important non-Gaussian features.

To address this limitation, researchers have proposed various non-Gaussian statistics for studying cosmological fields. These include the three-point function or bispectrum~\cite{10.1046/j.1365-8711.2003.06868.x, 2023JCAP...10..028H, 10.1111/j.1365-2966.2004.07410.x}, peaks and void counts~\cite{2000ApJ...530L...1J, 2009ApJ...698L..33M, 2018MNRAS.474..712M, 2021MNRAS.506.1623H, 2022MNRAS.511.2075Z, 2023arXiv230810866M}, topological methods ~\cite{2012MNRAS.419..536M, 2012PhRvD..85j3513K, 2022OJAp....5E..13G, 2019JCAP...09..052F, 2021A&A...645A.123P, 2021A&A...648A..74H, 2022A&A...667A.125H}, one-point-like statistics~\cite{Patton:2016umg, 2023MNRAS.519.4856B, 2023MNRAS.526.5530A}, wavelet transform based statistics~\cite{2020MNRAS.499.5902C, 2021arXiv211201288C, 2022PhRvD.106j3509V, 2023arXiv231100036H, 2022mla..confE..40P, DES:2023qwe}, and machine learning assisted methods~\cite{Fluri2018, Ribli2019, 2018PhRvD..97j3515G, 2019PhRvD.100f3514F, 2019NatAs...3...93R, 2020PhRvD.102l3506M, 2021MNRAS.501..954J, 2022MNRAS.511.1518L, Zhong:2024qpf, DES:2024xij}. For a forecast with the settings of the Euclid Mission, see Ref.~\cite{2023A&A...675A.120E}.

However, synthetic checks for these non-Gaussian statistics are not as well-developed as those for two-point statistics. One method of calibrating and validating results is through large-scale simulations of weak lensing maps, using either $N$-body or hydrodynamical simulations. These simulations are computationally expensive, making it infeasible to validate results over a large space of cosmological and nuisance parameters. The task of defining a covariance matrix of the summary statistics requires even more synthetic realizations of the field; even the covariance of the 2-point statistic is a non-Gaussian statistic. The desire for "field-level inference" (FLI) of cosmological parameters---marginalizing over the full mass field, eschewing summary statistics---is even more limited by the availability of synthetic maps that are fast, accurate, and differentiable. Alternative quasi-$N$-body simulations that require fewer computational resources exist~\cite{Scoccimarro:2001cj, Kitaura:2013cwa, White:2013psd, Tassev:2013pn, Izard:2015dja, Feng:2016yqz}, but they often lack accuracy at the smallest scales. Consequently, $N$-body or hydrodynamical simulations remain essential for validating results in future cosmological surveys.

Another alternative is to numerically produce maps that generate correct statistics, without calculating forces from physical models. One such approximation is the LogNormal (LN) field (see Sec.~\ref{sec:lognormal} for a short review). A LogNormal map $y$ is created by transforming a Gaussian random field $x$ by taking the exponential: $y_{\mathrm{LN}} = e^{x}.$ The name LogNormal comes from the fact that the 1-point distribution function (PDF) of the log of the variable $y$ follows a normal distribution. LogNormal maps have been used to model both overdensity field $\delta$ and weak lensing convergence $\kappa$ for over a decade~\cite{10.1093/mnras/248.1.1, Taruya:2002vy, 2011A&A...536A..85H, Xavier:2016elr}. In the context of the galaxy field, the exponential transformation also has the benefit of regularizing the bias terms. Under reasonable noise level and accuracy requirements, LogNormal models have been found to produce correct covariance matrices and bispectra~\cite{Hall:2022das}. However, the skewness of a LogNormal model is insufficient to represent the non-linear features of maps driven by gravitational evolution, leading to inaccuracies in higher-order statistics~\cite{Piras:2022dgt}.  LogNormal models can, by design, precisely emulate all 2-point functions of an $N$-body (or other simulated) map, and their PDF is a better approximation to the Universe's mass PDF than a purely Gaussian generator yields.  They have been used extensively in current-era WL survey analyses, particularly the \texttt{FLASK}~\footnote{\url{https://github.com/hsxavier/flask}\href{https://github.com/hsxavier/flask}{}} and \texttt{GLASS} code~\footnote{\url{https://github.com/glass-dev/glass}\href{https://github.com/glass-dev/glass}{}}, as surrogates for $N$-body models in applications where many realizations are needed. This is particularly useful in developmental stages, and in estimating covariances for summary statistics when analytic methods are insufficient \citep{DES:2017qwj}. Creating a LogNormal simulation requires specifying the 2 parameters of the transformation function in $y=G^{-1}(x)$ from the Gaussian field $x$ to the target field (mass overdensity $\delta$ or convergence $\kappa$), plus the specification of the power spectrum $P_x(k)$ of the Gaussian field.  The parameters of $G^{-1}$ and $P_x$ must be known as functions of $z$ and of cosmological parameters.

In this paper, we generalize the LN approach to "general point transforms of Gaussians" (GPTG) by allowing the transformation function $G^{-1}$ to be more complex than the exponential function to better approximate the $N$-body PDF. We then test whether the resultant maps can more accurately reproduce the higher-order statistics of the $N$-body fields that they emulate. We also investigate the limitations of this method, specifically regarding spatial/angular resolution of the maps and shape noise levels. While it is theoretically possible to find arbitrarily accurate functional forms to match the probability distribution function (PDF) of $N$-body simulations, there is a trade-off: more complex functions (more degrees of freedom) become more challenging to emulate across different cosmologies without overfitting or bias. To address this challenge, we employ symbolic regression algorithms to help us select functions that balance accuracy and complexity. As demonstrated in Sec.~\ref{sec:finding_inv_G_func}, we can achieve this goal with only 3 to 5 free parameters.

Generative models in weak lensing and galaxy field simulations have seen significant advancements in recent years, largely due to machine learning techniques. These include Generative Adversarial Networks (GAN)-based~\cite{2019ComAC...6....1M, Perraudin:2020gig, Tamosiunas:2020rvw, Yiu:2021pga, Shirasaki:2023nnk, Boruah:2024rgr, Bhambra:2024uvr}, Diffusion based model~\cite{Remy:2022ixn, Mudur:2023smm}, and Normalizing Flows~\cite{Dai:2022dso}. The point-transformation approach, if sufficiently accurate, can be considered an analytical generative model with very few free parameters that is very easy to train and fast to realize. While machine learning-based generative models can reproduce maps that are almost identical to $N$-body maps down to the smallest scales, most are limited to a single cosmology due to the difficulty of training over a large parameter space or robustly conditioning the same neural network model on various cosmologies (for a recent approach that enables cosmological dependence, see Ref.~\cite{Bhambra:2024uvr}). In contrast, the point transformation approach is easily generalizable to different cosmologies due to its limited number of parameters. This model offers several useful applications:
\begin{itemize}

\item Validation and blind testing for higher-order statistics: As various summary statistics and machine learning methods provide increasingly tighter constraints on cosmological parameters, the importance of validation and blind testing grows. Blind tests with unknown input cosmological parameters are one of the best methods to eliminate potential confirmation bias and compare different Bayesian inference approaches. See Refs.\cite{Nishimichi:2020tvu} and \cite{Beyond-2pt:2024mqz} for applications in galaxy clustering. The method outlined in this work serves as a potential toy model for conducting blind tests to compare different inference methods, over a large parameter space if needed.

\item Estimating covariance matrices: When analytical formulas for calculating covariance matrices for certain summary statistics are unavailable, running numerous simulations is often necessary. The point transformation method can generate different realizations and provide covariance matrix estimates, similar to the LogNormal case. If the generated maps produce sufficiently accurate summary statistics, they could, in principle, be used to estimate their covariance. Although using mock simulations in precision cosmology might be subject to unknown biases, it could still be valuable for testing assumptions and other observational effects~\cite{Shirasaki:2019gya}. A rough covariance matrix estimation could also help data compression methods~\cite{Park:2024pxd}.

\item Active sampling: With the need to sample a large parameter space in future surveys, active sampling—where we generate denser simulations in specific regions—can be crucial. Mock simulations like those presented in this work can provide rough estimations of posterior regions, allowing for targeted $N$-body or hydrodynamical simulations in those parameter spaces. This approach could help overcome computational challenges posed by the curse of dimensionality.

\item Augment and validate Machine Learning (ML) approaches: As ML methods are increasingly applied to cosmological inference~\cite{Fluri2018, Ribli2019, 2018PhRvD..97j3515G, 2019PhRvD.100f3514F, 2019NatAs...3...93R, 2020PhRvD.102l3506M, 2021MNRAS.501..954J, 2022MNRAS.511.1518L, Zhong:2024qpf, DES:2024xij}, the training stage of neural networks requires substantial input data, especially for sophisticated models like Transformers. Lognormal maps have already been used in ML investigations, primarily due to their ability to generate numerous maps or even provide new realizations on the fly~\cite{Akhmetzhanova:2023hiy, vonWietersheim-Kramsta:2024cks}. A more accurate model that maintains the advantages of LogNormal maps could, in principle, facilitate ML investigations in general.

\item Full Bayesian field-level inference: Field-level inference, where sampling occurs at the pixel level, is a promising direction for lossless cosmological parameter inference~\cite{2010MNRAS.407...29J, 2013MNRAS.432..894J, Porqueres:2021clw, Boruah:2023fph, Zhou:2023ezg}. Most field-level inference pipelines rely on forward models, with popular choices including Gaussian, LogNormal, or Lagrangian perturbation theory (LPT) models. An improvement over LogNormal models could benefit multiple aspects of field-level inference, including faster convergence and less biased or overconfident results.

\end{itemize}

The paper is outlined as follows. In Sec.~\ref{sec:methodology}, we discuss the details of the method including how we search different equations with symbolic regression, and the numerical methods for generating the Gaussian power spectrum. In Sec.~\ref{sec:validation_w_statistics} we test the model with different summary statistics including moments, wavelet, Minkowski functionals, and peak counts. In Sec.\ref{sec:applications_examples}, we demonstrate how it can be generalized to different cosmologies, how to estimate covariance matrices and test the limiting resolutions. We discuss potential future improvements and summarize conclusions in Sec.~\ref{sec:conclusions}.

\section{Methodology}\label{sec:methodology}

\subsection{Weak Lensing Simulations}\label{sec:simulations}

In this study, we explore the potential for directly modeling the weak lensing convergence field $\kappa$ across multiple tomographic bins. This methodology could also be applied to density shells, $\delta^\Sigma$, which can later be converted into the convergence field through convolution with lensing kernels. This approach is analogous to choosing between using a LogNormal distribution to model either weak lensing maps directly or multiple mass shells, each offering distinct advantages. For a detailed comparison, see Ref.~\cite{Xavier:2016elr}.

Our map-making procedure from $N$-body simulations is similar to those used in Refs~\cite{y3-massmapping, DES:2023qwe, DES:2024xij}. The $N$-body simulation suite we used in this work is the dark-matter-only \texttt{DarkGridV1}~\cite{Zuercher2021,2022MNRAS.511.2075Z} that spans a cosmological space of $\Omega_{\rm m} $ and $\sigma_8$ with 57 different samples. Each cosmology is represented by five independent full-sky simulations. The simulations utilize the \textsc{PKDGRAV3} code \citep{potter2017pkdgrav3}, producing particle number maps at 100 different redshifts, ranging from $z=49$ to $z=0$. These particle number maps are then converted to noiseless convergence maps using the Born approximation~\cite{Fosalba2015}. We subsequently generate integrated DES-Y3-like shear maps by weighting the shear maps as a function of redshift according to the DES-Y3 redshift distributions~\cite{y3-sompz}. Each resulting $\kappa$ map consist of four highly correlated channels, representing the lensing effect on four different redshift bins of source galaxies. For most of this work, we consider the fiducial cosmology at $ \Omega_\mathrm{m}=0.26$ and $\sigma_8 = 0.84$. The resolution of the available maps is 6.9 arcmin or $\rm N_{side}=512$ in \texttt{HEALPIX}. We do not add further systematics and rely on maps without any intrinsic alignment contributions.

We further test our method at higher resolutions in Sec.\ref{sec:limits_of_point_transform} using the Takahashi17 simulations \cite{Takahashi:2017hjr}. These simulations provide multiple ray-tracing convergence maps at different redshift shells with a depth of 150 $h^{-1}$ Mpc. We generate the same four-channel maps by taking the weighted average of the convergence field at each shell, $\kappa^{j}_{\mathrm{shell}}$, according to the equation:
\begin{equation}
\kappa^i=\frac{\sum_{j=1}^{N_{\text {shells }}} n^i\left(z_j\right) \kappa_{\text {shell }}^j}{\sum_{j=1}^{N_{\text {shells }}} n^i\left(z_j\right)},
\end{equation}
where $n^i\left(z_j\right)$ is the binned DES-Y3 redshift distribution as in \texttt{DarkGridV1}. The use of two different simulation suites is motivated by the fact that \texttt{DarkGridV1} explores different cosmological parameters, but only at low resolution, whereas Takahashi17 offers high-resolution maps but at a single fixed cosmology. The differences between the two simulations' computational methods are unimportant, as we will see that the accuracy of the GPTG emulation is primarily a function of the angular resolution of the map we are attempting to emulate.

\subsection{One-point transformation and Gaussianization function}

\begin{figure}[t]
\centering
\includegraphics[width=0.99\columnwidth]{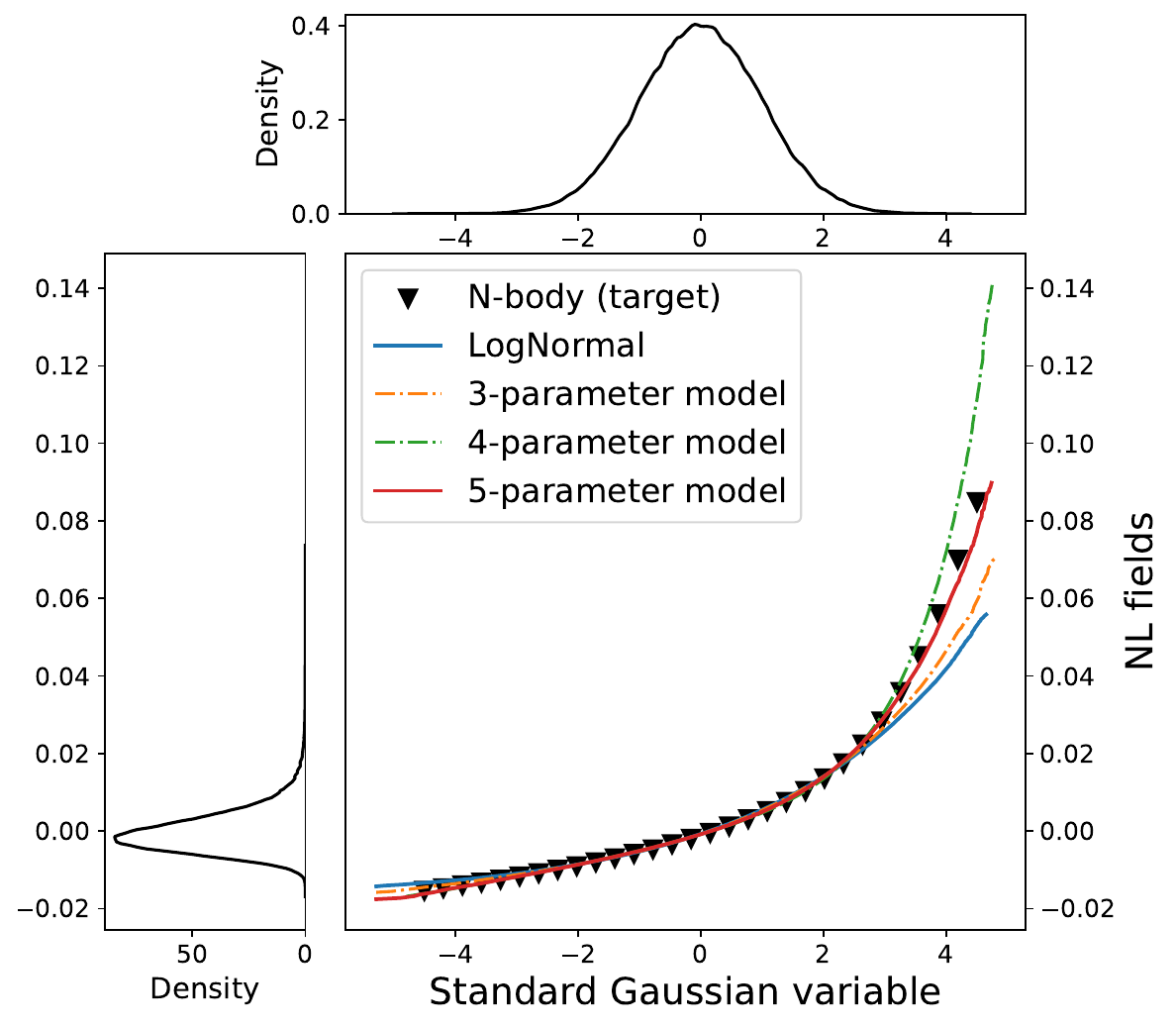}

\caption{We plot the inverse of the Gaussianization function $G^{-1}$ defined in eq.~\ref{eq:def_G_function}. Black points are from the $N$-body simulations of the $\kappa$ field with a resolution of 6.9 arcmin.  $G^{-1}$ is the function that maps each point of a standard Gaussian field (value on $x$ axis) to the non-Gaussian $\kappa$ field (values on $y$ axis). The plot shows the third tomographic bin as an example; the other bins are similar. The colored lines plot several analytic $G^{-1}$ functions examined in this work. The generalized functions proposed here are more flexible in describing both tails than the simple exponential used in the LogNormal case.
} 
\label{fig:G_curve} 
\end{figure}
A field $y_{\mathrm{sim}}$ that has a non-Gaussian PDF can be transformed to another field that has a standard Gaussian PDF (mean 0 and standard deviation 1) via a local transformation at each point, or a \textit{Gaussianization} function:
\begin{equation}\label{eq:def_G_function}
\begin{aligned}
    G(y_{\mathrm{sim}})&=\operatorname{ppf}^{\mathrm{norm}} \left(\operatorname{cdf}(y_{\mathrm{sim}}) \right)\\
&=\sqrt{2}\operatorname{erf}^{-1}\left(2 \operatorname{cdf}(y_{\mathrm{sim}})-1\right) ,
\end{aligned}
\end{equation}
where we take the cumulative distribution function (CDF) of the non-linear fields $y_{\mathrm{sim}}$ and match the quantiles to the standard normal distribution. We do this by taking the inverse of the CDF, or percent-point function (PPF), which is called the probit function for the normal distribution and can in turn be obtained from the error function, for which several approximation methods exist~\cite{abramowitz_stegun}. In practice, we use the \texttt{norm} distribution class in \texttt{Scipy} to calculate the Gaussianization function. The inverse $G^{-1}$ of the Gaussianization function estimated from one of the $N$-body maps is plotted in Fig.~\ref{fig:G_curve}, along with the LogNormal and other analytic approximations explored in this paper.

The \textit{point-transformed Gaussia} method, or \textit{inverse-Gaussianization} as called in Ref~\cite{Yu:2016qoq, 2011PhRvD..84b3523Y}, to generate different realizations of a large-scale structure field can be summarized as follows. We first find the $G$ function according to eq.~\ref{eq:def_G_function} and get the Gaussianized field $G(y_\mathrm{sim})$. Ignoring the residual non-Gaussianity hidden in the higher order correlation functions (e.g. due to filamentary structures), the power spectrum of the Gaussianlized field captures all the information. We can generate different realizations of the Gaussian field, as many as we want, by drawing different Fourier modes according to the power spectrum. We then apply the inverse of the Gaussianization function as a local transform and get a different realization of the non-linear field $y_{\mathrm{gen}}$. 

The accuracy of the above method has been tested in~\cite{2009ApJ...698L..90N, Yu:2016qoq, 2011PhRvD..84b3523Y, Qin:2020old, 2020ApJ...897...14C} with summary statistics such as the power spectrum and bispectrum. It was shown that the generated maps in general lose the filamentary structure but produce accurate summary statistics. At the same time, since the summary statistics are recovered using such a method, it means the Gaussianization function itself is also a good summary statistic. In essence, it is the same as using the PDF, but the Gaussianization function provides an alternative description with a much smaller data vector. This idea has previously been tested in the context of both weak lensing maps and galaxy maps. Although improvements have been shown in optimistic simulations, little gain is obtained with the presence of realistic noise and systematics~\cite{2010ApJ...708L...9S, 2012PhRvD..86b3515Y, 2011MNRAS.418..145J}.

The proposed idea in this paper is to analytically parameterize the inverse of the Gaussianization function $G^{-1}$ with a simple functional form, which can also be regarded as a simple structure formation function. The key difference from using a simple logarithmic to Gaussianize the field is that the function does not necessarily have a closed-form inverse function. Note that an analytical approach such as the Box-Cox transform~\cite{b6d53586-2890-3ac6-bec5-c3cfdcb64879, 2011MNRAS.418..145J} has been applied to cosmology, where a power-law parametric form of the Gaussianization is used. However, parameterizing the inverse of the Gaussian function is more straightforward in the application of forward modeling or generating new maps. We address the difficulties of getting the corresponding power spectrum in Sec.~\ref{sec:power_spectrum_for_generation}.

\subsection{LogNormal Fields}\label{sec:lognormal}

The exact definition of the LN field differs in the literature depending on the use case. Here we use a 2-parameter model (sometimes called the shifted lognormal):
\begin{equation}\label{eq:def_LogNormal}
    y_{\mathrm{LN}} = \beta \mathrm{e}^{\alpha x-\alpha^2/2} - \beta ,
\end{equation}
where $x$ is a standard normal variable with mean 0 and variance 1; $\alpha$ and $\beta$ are two free parameters to fit the target distribution. It naturally gives zero mean $\left\langle y_{\mathrm{LN}}\right\rangle = 0$ since $\left\langle \exp{[\alpha x - \alpha^2/2]} \right\rangle = 1$. The $\beta$ is called the "shift parameter" in FLASK and GLASS~\cite{Xavier:2016elr, Tessore:2023zyk}, and it improves the fit to the PDF of $N$-body PDF $\kappa$ maps. It is more common to fix the shift parameter to $\beta=1$ when modeling overdensity fields due to the constraint that the density cannot be lower than 0.

One notable advantage of the LogNormal variable is that it has analytical correlation functions at all orders. The relation between the two-point correlation functions $\xi^{ij}$ of the LogNormal field and that of the underlying Gaussian field is:
\begin{equation}\label{eq:LN_corr}
\xi_{\mathrm{LN} }^{i j} \equiv\left\langle y_{\mathrm{LN}}^i y_{\mathrm{LN}}^j\right\rangle = \beta_i \beta_j\left(e^{\xi_{\mathrm{G}}^{i j}}-1\right)
\end{equation}
where $\xi_{\mathrm{G}}^{i j}$ is the two-point correlation function of the Gaussianized field (auto or cross). This allows programs to use the theoretical calculation from Boltzmann codes to estimate the power spectrum, as detailed in Ref.~\cite{Tessore:2023zyk}.

\begin{table*}[t]
\centering
\begin{tabular}{|l|c|c|}
\hline
Name & Point Transform Function & Features \\ \hline
LN ($G^{\mathrm{inv}}_2$)& $\beta \exp(\alpha x-\alpha^2/2) - \beta$ &   \makecell{The PDF and relation of correlation functions \\ to the Gaussian field are analytical. 
}\\ \hline

$G^{\mathrm{inv}}_3$ & $n\left(\mathrm{e}^{a x - a^2/2} + bx + c \right) - 1$ & \makecell{Best function suggested by Symbolic Regression. \\ Fits individual tomographic bins, \\ but failed to model cross-bin correlations.} \\ \hline

$G^{\mathrm{inv}}_4$ & $n \mathrm{e}^{a_1 x-a_1^2 / 2}\left(1+\mathrm{e}^{\left(x-x_0\right) t}\right)^{(a_2-a_1)/t}-1$ & \makecell{Functional form motivated by the limiting \\ behavior of Gaussianization curve, previously tested \\ on density shells with Quijote simulations \cite{Porth:2021tva}.}\\ \hline

$G^{\mathrm{inv}}_5$ & $n \left(\mathrm{e}^{a_1 x-a_1^2 / 2 } + b x \right) \left(1+\mathrm{e}^{\left(x-x_0\right) t}\right)^{(a_2-a_1)/t}-1$ & \makecell{Functional form that combines  the two functions above. \\ The two tails are flexible enough to accurately\\   model the $\kappa$ maps with different cosmology.} \\ \hline

\end{tabular}
\caption{Summary of point transform functions studied in this paper. The $n$ factor ensures that the transformed fields have zero mean. In the case of our 3-parameter model $G^{\mathrm{inv}}_3$, $n=1/(1+c)$ while in other cases $n$ is numerically estimated. Note that 2 parameters of the 5-parameter model, $x_0$ and $t$, are insensitive to cosmological parameters at a given tomographic bin. 
For more details, see Sec.~\ref{sec:vary_cosmology}.}
\label{table:formulas}
\end{table*}

\subsection{Finding the inverse-Gaussianization functions}\label{sec:finding_inv_G_func}

In this section, we summarize the process of finding the inverse-Gaussianization function that best fits the data. The definition of "best fit" is subject to the final goals and involves several considerations. In theory, with sufficient parameters and flexible functions, the fit to the inverse-Gaussianization function can be made arbitrarily accurate. However, this approach may hinder generalization to different cosmologies, as the best-fit parameters might vary excessively and not smoothly with cosmological parameters. The definition of accuracy should be based on both the PDF difference and other summary statistics. We aim to fit both tails as an improvement over the LogNormal model, while still prioritizing the central region where most data points lie. For these reasons, we seek to combine numerical methods and human expectations to evaluate the goodness of fit to the inverse-Gaussianization functions. The functions we considered are summarized in Table~\ref{table:formulas}.

We use Symbolic Regression (SR) to help us finding the best functions in this work. SR algorithms have been developed in recent years to automatically search for the best-fitting formula over a large functional space. Examples of SR use in astrophysics include~\cite{Bartlett:2022kyi, 10.1093/mnras/stae803, Alestas:2022gcg, 2022MNRAS.515.2733D, 2023MLS&T...4d5002L, Bartlett:2024jes}, where it has been applied to rediscover physical laws, find relations in complex data sets, or correct existing fitting formulae. SR algorithms typically balance complexity and accuracy using specific ranking criteria. SR falls within the machine learning domain, focusing on interpretability and finding useful latent space representations.

In this work, we explore SR using the package \texttt{PySR}~\cite{cranmerInterpretableMachineLearning2023}. After trying various settings and normalizations, we consistently found that an expression with exponential plus linear terms emerges as the winner: $y=a\mathrm{e}^{bx}+cx+d$, which we call the 3-parameter model. This form also ranked highly when fitting the left or right tails separately.  The function has 3 free parameters once we enforce normalization to zero mean.

We investigate two other analytic forms with successively more free parameters.  The 4-parameter model specified in Table~\ref{table:formulas}, previously used in Ref.~\cite{Porth:2021tva} as "DoubleLog", is designed to interpolate between two straight lines with different slopes ($a_1$ and $a_2$) in $x$-vs-$\log y$ space, with parameter $t$ controlling the transition speed and $x_0$ determining the transition point. With inspiration from the SR results, we investigated the final 5-parameter form in the Table, which adds a linear term to one of the exponential factors in the DoubleLog transform to yield 5 free parameters. We denote these three models as $G^{\mathrm{inv}}_n$ (inverse-Gaussian function with $n$ parameters) and summarize their fits in Fig.~\ref{fig:G_curve}. As shown in Sec.~\ref{sec:validation_w_statistics}, the 5-parameter model has enough freedom to fit the PDF well across different cosmologies. However, we do not expect the exact form of the 5-parameter model to matter so much. For example, an expansion form like $a\mathrm{e}^{bx}+cx+dx^2+ex^3+f$ could also serve as a 5-parameter model.

\subsection{Power Spectrum for Map Generation}\label{sec:power_spectrum_for_generation}
The relation between the power spectrum of the Gaussianized field and the transformed field has a closed form only in some special cases, including LN. However, one can always obtain a good approximation by Taylor-expanding either the Gaussianizing function or the inverse-Gaussianizing function. In theory, the relation in either direction  depends on all the higher-order N-point functions. The relation is even more complicated in the presence of noise~\cite{Crocce:2005xz, 2009ApJ...698L..90N, 2011MNRAS.418..145J}.  

We can derive an angular power spectrum $C_\ell^G$ that can be used to generate the Gaussian field $x$ such that the transformed non-Gaussian field $y=G^{-1}(x)$ has precisely any desired power spectrum $C_\ell^{NG},$ for any monotonic $G^{-1}.$
 A similar procedure was used in Ref.~\cite{Tessore:2023zyk}.\footnote{see Section.~\ref{sec:validation_w_statistics} therein for a detailed comparison in the LogNormal case} We summarize the basic procedures as follows.
\begin{equation}\label{eq:Cl_procedure}
    C_\ell^{\mathrm{NL}} \xrightarrow{\text{\textcircled{1}}} \xi^{\mathrm{NL}} (\theta) \xrightarrow{\text{\textcircled{2}}} \xi^G (\theta) \xrightarrow{\text{\textcircled{3}}} C_\ell^{\mathrm{G}}
\end{equation}
The matching criterion is expressed for the correlation functions $\xi$ rather than the power spectra $C_\ell,$ so we start by Fourier-transforming the target angular power spectrum of the $N$-body simulation maps $C_\ell^{\mathrm{NL}}$ into the two-point correlation function $\xi^{\mathrm{NL}} (\theta.)$  Step 2 is to find the value of $\xi^G(\theta),$ the correlation function of the Gaussianized field, which will yield the desired $\xi^{\rm NL}(\theta)$. The condition for this equality is:
\begin{equation}\label{eq:xi_pt_relation}
\begin{aligned}
    \xi^{\mathrm{GPTG} }_{i j} &\equiv\left\langle G^{-1}_i(x_i) G^{-1}_j(x_j)\right\rangle \\
    & = \iint \mathrm{d}x_i \mathrm{d}x_j G^{-1}_i(x_i) G^{-1}_j(x_j) \mathcal{N}\left(0 ,
    \begin{pmatrix}
    1 & \xi^{G}_{ij}\\
    \xi^{G}_{ij} & 1
    \end{pmatrix}\right) ,
\end{aligned}
\end{equation}
In this equation, the indices $ij$ label the different bins of a tomographic $\kappa$ or $\delta$ field, and  $G^{-1}_i$ and $G^{-1}_j$ are assumed to be known. By definition, $x_i$ and $x_j$ are part of a Gaussian field and must have a normal joint distribution with unit variance per dimension.  Eq.~\ref{eq:xi_pt_relation} therefore offers an implicit solution for $\xi^G_{ij}$ at every value of $\theta.$  This can be found via numerical integration, although in some cases the integral is analytic (such as LogNormal, eq.~\ref{eq:LN_corr}). Step 3 of the algorithm is to transform the $\xi^G(\theta)$ functions back into a power spectrum $C_\ell^G$ for the Gaussianized variable(s) $x.$ We use the \texttt{Healpy} routine \texttt{synfast} to generate a full-sky Gaussian field with the input power spectrum $C_\ell^{\mathrm{G}},$ then apply the $G^{-1}$ transform to the output. Note that the method is applicable equally well to a tomographic series of maps as to a single map, using the cross-tomographic-bin power spectra. The resulting multi-channel maps generated will reproduce the correct cross-bin non-linear power spectra. 

More precisely, we perform step 2 by creating a look-up table of Eq.~\ref{eq:xi_pt_relation}, then use linear interpolationto find the $\xi_{G}$ at each $r$ that gives $\xi_{\mathrm{GPTG} }^{i j}$ to match that of the non-linear field $\xi_{\mathrm{NL}}$. We tested other interpolation methods and found no difference. However, to account for rare points that lie beyond the interpolation range, we found that linear extrapolation gives the most stable results. We thus guarantee that the two maps, no matter how good the PDF fit is, will always have the same two-point statistics. This is also why the LogNormal fields as in Ref.~\cite{Tessore:2023zyk} have the correct second moments despite the inaccurate PDF. We checked the second-order information including power spectrum and second moments $\left\langle\hat{\kappa}_{\theta_0}^2\right\rangle$ and all the maps agree with the $N$-body maps (including the LogNormal ones).

A numerical challenge remains in transforming between the angular power spectrum and the two-point correlation function (steps 1 and 3). Given isotropy, the two are related by:
\begin{equation}\label{eq:Cl2xi}
\xi(\theta)=\sum_{l=0}^{\infty} \frac{2 l+1}{4 \pi} C_l P_l(\cos \theta),
\end{equation}
and
\begin{equation}\label{eq:xi2Cl}
C_l=2 \pi \int_{-1}^{1} \xi(\theta) P_l(\cos \theta)  \mathrm{d} \cos (\theta),
\end{equation}
where $P_l$ is the $l$-th order Legendre polynomial. Obviously, we could not sum to infinity. We also find that using the direct definition above the generated maps suffer from numerical instabilities in the high $l$ region. Instead, we use the algorithm proposed in Ref.~\cite{doi:10.1137/0912009}, which is also used previously in the LogNormal generator of~\cite{Tessore:2023zyk}. The basic idea is to relate the Legendre expansion of a function:
\begin{equation}
f(\theta)=\sum_{l=0}^{n-1} a_l P_l(\cos \theta)
\end{equation}
to that of the Fourier cosine expansion
\begin{equation}
f(\theta)=\sum_{k=0}^{n-1} b_k \cos (k \theta) .
\end{equation}
The relation of the coefficients $a_l$ and $b_k$ is linear with a matrix of Gamma functions. One can thus efficiently calculate $f(\theta)$ expansion by Discrete Cosine Transform (DCT) and recursively get the Legendre expansion coefficients. We use the algorithm implemented in \texttt{cltools}~\footnote{\url{https://github.com/cltools/cltools}\href{https://github.com/cltools/cltools}{}}, and we refer there for a detailed explanation.

This slightly convoluted procedure ensures the point-transformed field has the same power spectrum as the non-linear field regardless of the fitting of the PDF. Another benefit of using the correlation function of the non-linear field rather than the Gaussianized field is to relate to theoretical code such as Boltzmann solver \texttt{CAMB}~\cite{2011ascl.soft02026L}.

\begin{table}[t]
\renewcommand{\arraystretch}{1.1}
\centering
\begin{tabular}{|c|c|c|}
\hline
Statistics           & Deviation of LogNormal     & Deviation of the 5-parameter \\ \hline

PDF z1      &   N/A               &   0.46  \\ \hline
PDF z2      &   N/A               &   0.32  \\ \hline
PDF z3      &   N/A               &   0.40  \\ \hline
PDF z4      &   N/A               &   0.47  \\ \hline
\multicolumn{3}{|l|}{\textbf{\textsc{}}}                \\ \hline
$\kappa_i^3$, $i=1$      &   0.71               &   0.33  \\ \hline
$\kappa_i^3$, $i=2$      &   1.12               &   0.28  \\ \hline
$\kappa_i^3$, $i=3$      &   1.26               &   0.41  \\ \hline
$\kappa_i^3$, $i=4$      &   1.23               &   1.11  \\ \hline

\multicolumn{3}{|l|}{\textbf{\textsc{}}}                \\ \hline
$\kappa_1*\kappa_i^2$, $i=2$      &   1.36               &   0.56  \\ \hline
$\kappa_1*\kappa_i^2$, $i=3$      &   1.54               &   0.48  \\ \hline
$\kappa_1*\kappa_i^2$, $i=4$      &   1.66               &   0.51  \\ \hline
\multicolumn{3}{|l|}{\textbf{\textsc{}}}                \\ \hline
ST1 z1      &   1.46               &   0.74  \\ \hline
ST1 z2      &   2.08               &   0.33  \\ \hline
ST1 z3      &   3.56               &   0.39  \\ \hline
ST1 z4      &   3.65               &   0.66  \\ \hline
\multicolumn{3}{|l|}{\textbf{\textsc{}}}                \\ \hline

ST2 z1      &   0.79               &   0.51  \\ \hline
ST2 z2      &   1.00               &   0.55  \\ \hline
ST2 z3      &   1.35               &   0.69  \\ \hline
ST2 z4      &   1.56               &   0.58  \\ \hline
\multicolumn{3}{|l|}{\textbf{\textsc{}}}                \\ \hline
MF z1      &   1.58               &   0.43  \\ \hline
MF z2      &   1.38               &   0.28  \\ \hline
MF z3      &   2.04               &   0.37  \\ \hline
MF z4      &   2.36               &   0.43  \\ \hline
\multicolumn{3}{|l|}{\textbf{\textsc{}}}                \\ \hline
Peaks z1      &   N/A               &   0.53  \\ \hline
Peaks z2      &   N/A               &   0.55  \\ \hline
Peaks z3      &   N/A               &   0.60  \\ \hline
Peaks z4      &   N/A               &   0.64  \\ \hline

\end{tabular}
\caption{Summary of LogNormal maps and 5-parameter maps performance. The deviation is defined as how many standard deviations away given LSST-Y10 error, namely $\Delta/\sigma_{\mathrm{LSST-Y10}}$ where $\Delta$ is the difference between LN or $G^{\mathrm{inv}}_5$ to the $N$-body maps. For a given data vector, we report the average value. For PDF and peak counts, the LogNormal model gives ill-defined values due to the lack of high tail values, we therefore list them as N/A. The 5-parameter model gives 2 to 5 times better agreement than LogNormal maps.} 
\label{table:diff_summary}
\end{table}
\section{Accuracy of non-Gaussian statistics of GPTG maps}\label{sec:validation_w_statistics}

\begin{figure*}
\centering
\includegraphics[width=0.99\columnwidth]{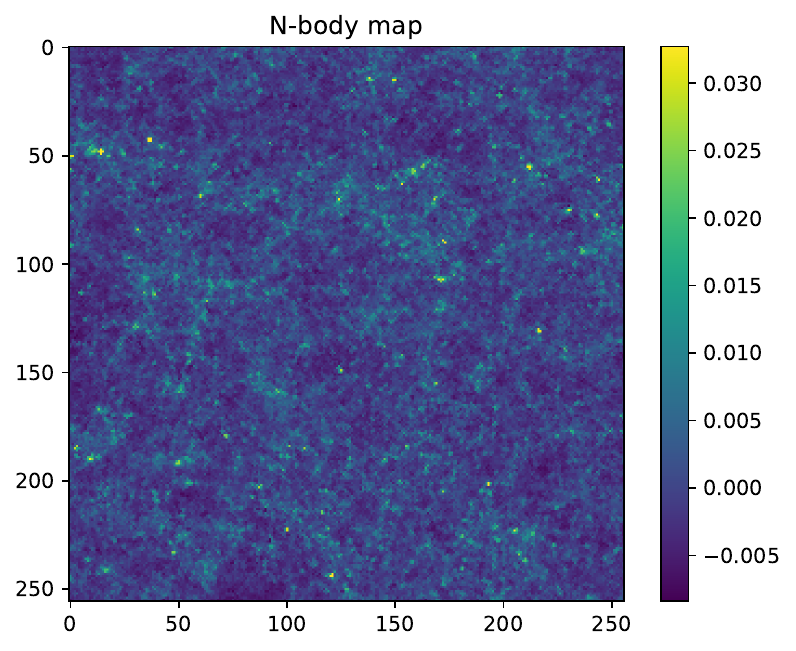}
\includegraphics[width=0.99\columnwidth]{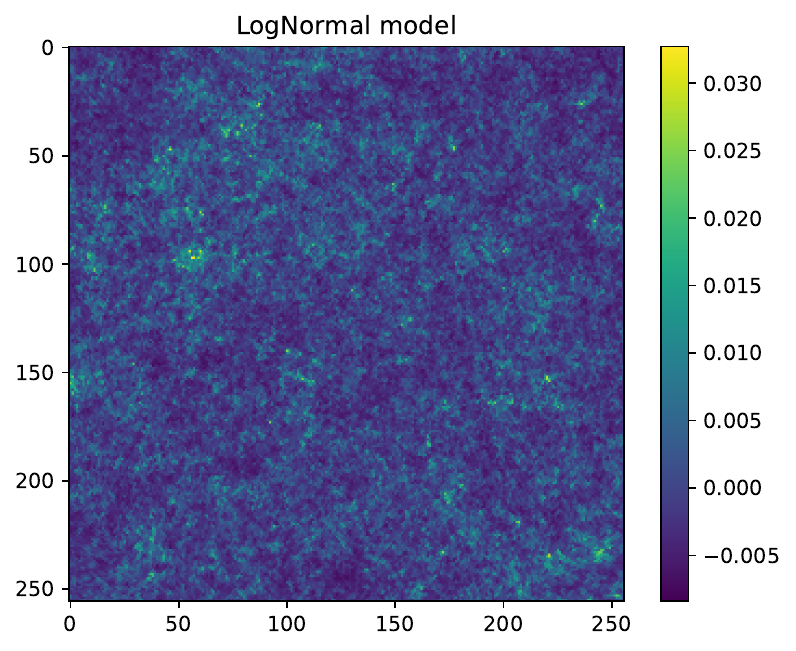}
\includegraphics[width=0.99\columnwidth]{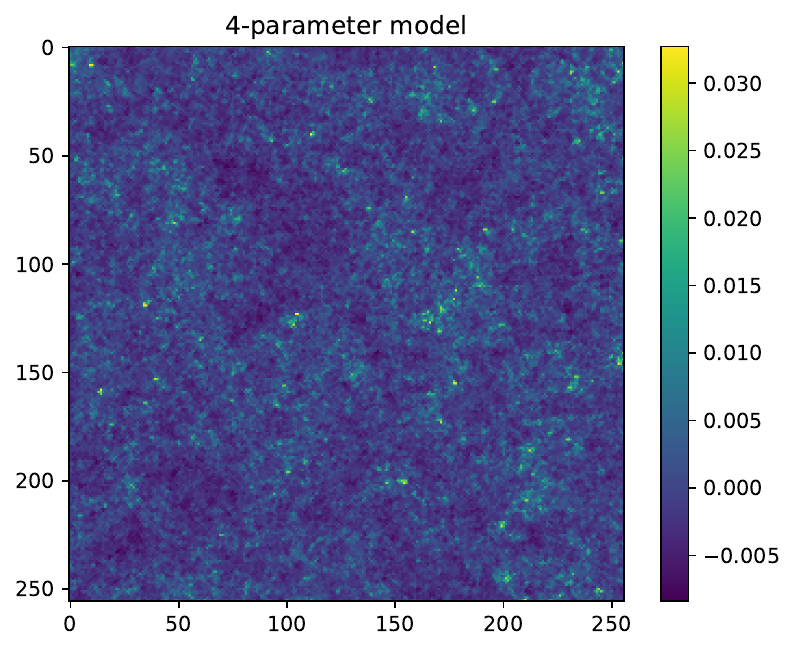}
\includegraphics[width=0.99\columnwidth]{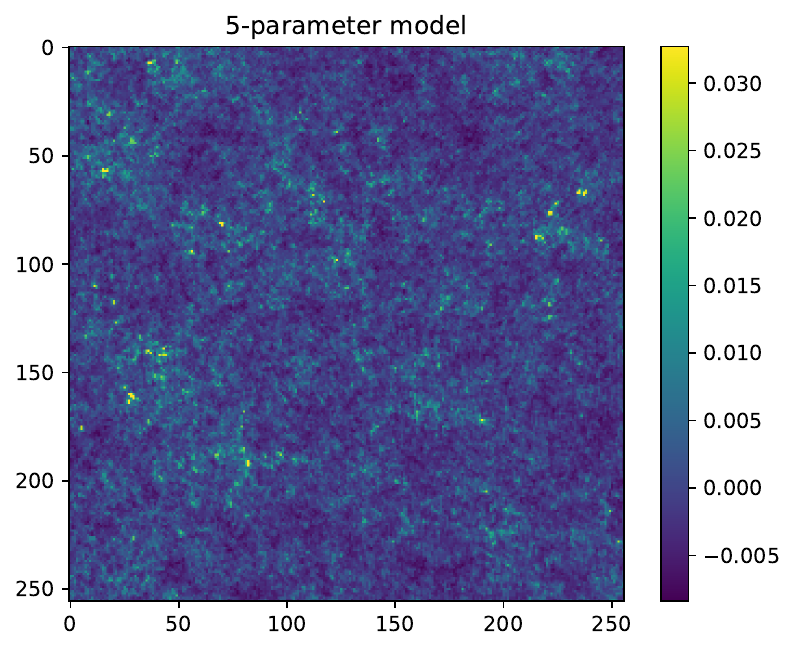}
\caption{Example noiseless patches of $N$-body maps and other GPTG maps. The color scale is set to be the same for a better visualization. The LogNormal case is found to be too smooth compared with other maps. Visually the 4-parameter and 5-parameter maps have more similar bright pixels as the $N$-body maps. The $N$-body map does show more filamentary structures that the point-transformed maps failed to produce. However, there are significantly fewer filaments in weak lensing convergence maps compared with overdensity maps, especially after shape noise. The main question we aim to answer in Sec.~\ref{sec:validation_w_statistics} is whether the difference is negligible for common summary statistics.
}
\label{fig:example_patch} 
\end{figure*}

In this section, we assess the fidelity of $\kappa$ maps generated with GPTG emulation of $N$-body outputs. For visual inspection, examples of sky patches with no shape noise are shown in Fig.~\ref{fig:example_patch}. The maps look very similar but we can still see some subtle differences between the generated maps and the $N$-body maps. The $N$-body maps have more local maxima and some moderate filamentary structures. These are indeed the limitations of the inverse-Gaussianization method, as also been pointed out in previous works~\cite{2009ApJ...698L..90N, Yu:2016qoq, 2011PhRvD..84b3523Y}. The purpose of this section is to verify whether these visual differences translate into significant deviations in other non-Gaussian statistics. The non-Gaussian summary statistics we tested in this work include third moments, scattering wavelet transform, Minkowski functionals, and peak counts. We do not show plots for power spectrum and second moments because all of these should be (and are) perfectly reproduced if eq.~\ref{eq:xi_pt_relation} has been enforced.

\begin{figure*}
\centering
\includegraphics[width=1.99\columnwidth]{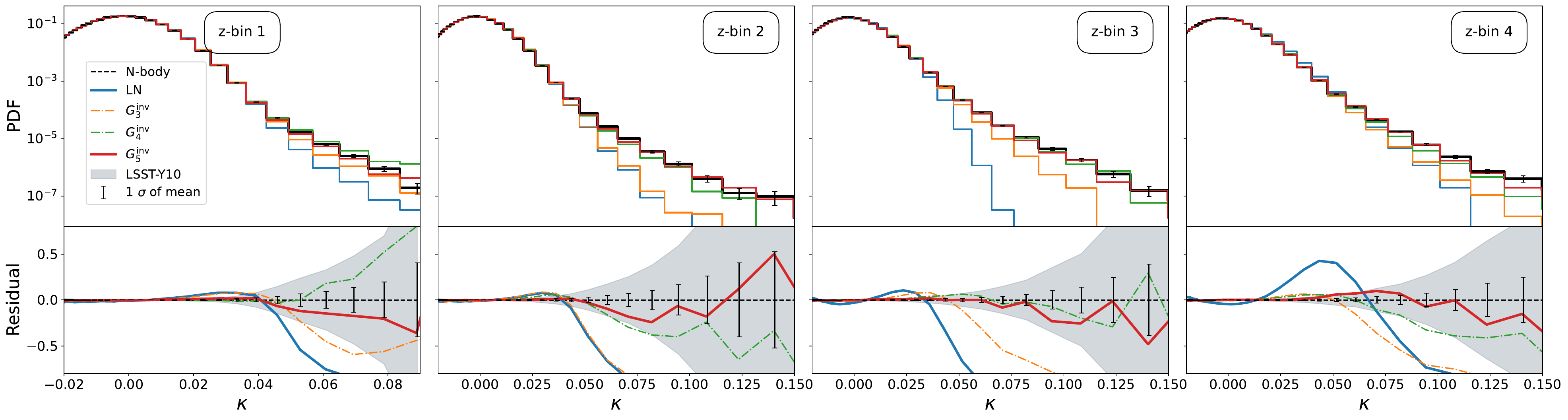}
\caption{The 1D probability distribution (PDF) of different maps using a log scale. Although in 2D the lowest bin is the most non-Gaussian, the higher bins have long tails of large values of $\kappa$ which makes the PDF challenging to model. The black error bars indicate the 1 $\sigma$ uncertainty in our simulations. The gray bands indicate the 1$\sigma$ level for an LSST-Y10-like survey.} 
\label{fig:PDF_comparison} 
\end{figure*}

We add shape noise as a Gaussian noise with standard deviation $\sigma_n$ following:
\begin{equation}
\sigma_n=\frac{\sigma_\epsilon}{\sqrt{n_g A_{p i x}}}
\end{equation}
where we use the shape noise $\sigma_\epsilon=0.26$ and galaxy density $n_g=27$ to represent a LSST-Y10-like observation. For the 4 tomographic bins used in this work, we neglect the galaxy distribution and divide $n_g=27$ by 4. The presence of shape noise will smear out the higher-order contributions such as filaments, and thus the generated noisy maps are expected to better reproduce statistics from the noisy $N$-body maps. We obtain the error bar of LSST-Y10-like data by assuming a sky coverage of $ 14400 \  \mathrm{deg}^2$. The corresponding PDFs are shown in Fig.~\ref{fig:PDF_comparison}. We summarize the results of different summary statistics in Tab.~\ref{table:diff_summary} by comparing the error of generated maps to the error bar of LSST-Y10.

\subsection{Third Moments}

\begin{figure*}
\centering
\includegraphics[width=1.99\columnwidth]{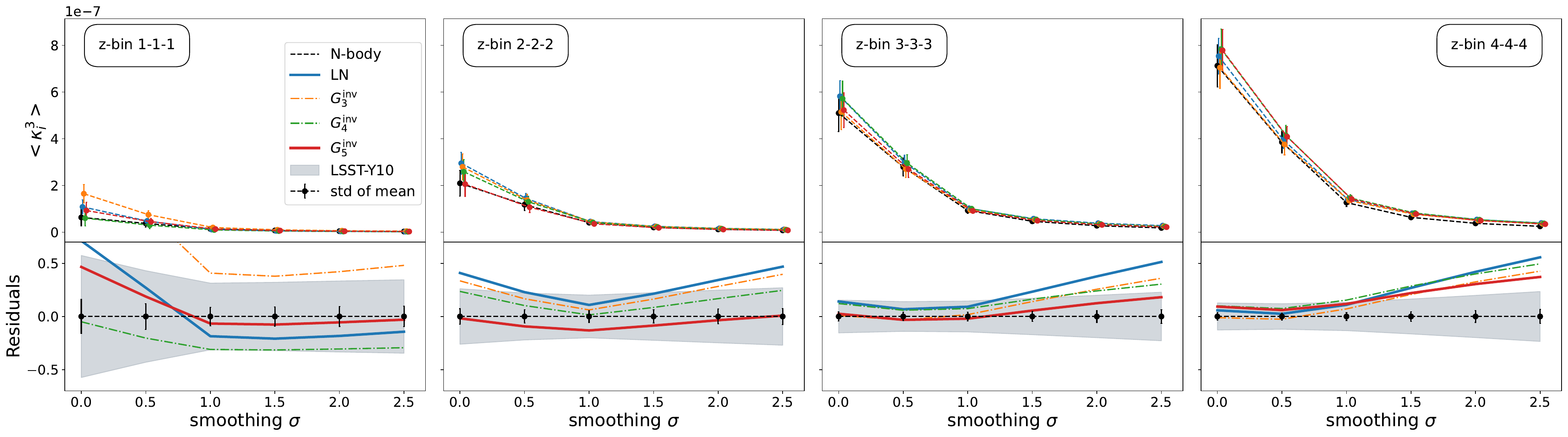}
\includegraphics[width=1.99\columnwidth]{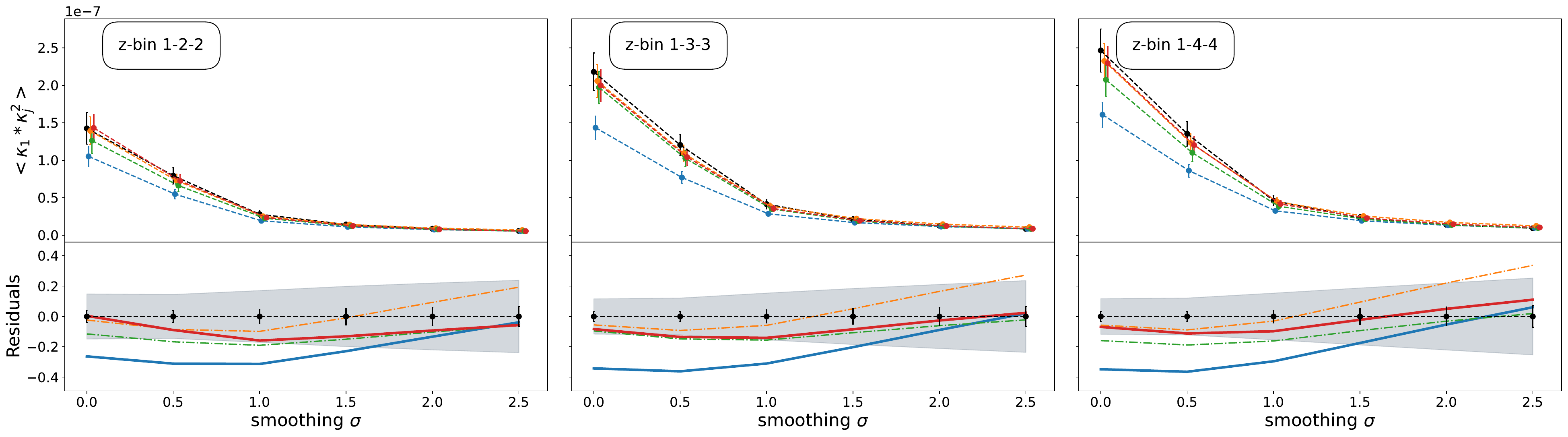}
\caption{Third moments of the generated maps compared with the $N$-body maps. Since we ensured that  the cross-bin correlations and power spectra are correct, the four tomographic bin maps are correlated and give the correct third moment across redshift bins, as shown in the lower panels. Note that the points in the upper pair of panels are shifted slightly along the x-direction for better visualization.} 
\label{fig:third_moment_comparison_tomo} 
\end{figure*}

The first kind of summary statistics we tested is the third moment~\cite{DES:2021lsy, DES:2022oqz, DES:2023qwe}. It is the spatial average of the cube of the field $\left\langle\hat{\kappa}_{\theta_0}^3\right\rangle$. Here $\theta_0$ denotes the scales of smoothing kernels applied to the maps before taking the cube; typically several will be used, yielding scale dependence of the 3-point function.  Note that when there is no smoothing of the maps beyond the pixel scale at which the GPTG is drawn, this statistic is essentially a test of the accuracy of the $G^{-1}$ functions in reproducing the $N$-body PDF.  Fitting the pixel-scale PDF does not, however, guarantee the accuracy of moments after smoothing to larger scales, making this a meaningful test of non-Gaussian structure.

We calculate the moments using Gaussian filters with standard deviations between $0$ and $2.5$ (pixel units, 6.9 arcmin here). The results are shown in Fig.~\ref{fig:third_moment_comparison_tomo} for both auto and cross-bin cases. By modeling the PDF and power spectrum simultaneously, we achieve much better accuracy in the third moments compared to the LogNormal maps (which can only be made to match the power spectrum). Compared with the estimated uncertainty for the LSST-Y10 observation, the GPTG maps give mostly consistent results, whereas the LogNormal maps give biased results up to 50\%. See Table~\ref{table:diff_summary} for the exact comparison.

\subsection{Scattering Wavelet Transform}
The Scattering Wavelet Transform (SWT)~\cite{https://doi.org/10.1002/cpa.21413} is a mathematical well-defined summary statistic that has been shown to be remarkably powerful in many fields of cosmology~\cite{2020MNRAS.499.5902C, 2021arXiv211201288C, 2022PhRvD.106j3509V, 2023arXiv231100036H, 2022mla..confE..40P, Gatti2023}. SWT gives a set of hierarchical coefficients by convolution of the filter with the field once or twice and takes the absolute value:
\begin{equation}\label{eq:def_of_ST}
\begin{aligned}
ST_1\left(j_1\right) & \equiv\left\langle\left|\mathbf{\kappa} \star \psi(j_1)\right|\right\rangle \\
ST_2\left(j_1, j_2\right) & \equiv\left\langle|| \mathbf{\kappa} \star \psi(j_1)\left|\star \psi(j_2)\right|\right\rangle .
\end{aligned}
\end{equation}
where $j_1$ and $j_2$ are parameters that control the scale of the wavelet. Brackets denote both the spatial average and the average of different angles of the filter. In this work, we follow the choice of Ref~\cite{2020MNRAS.499.5902C}, and use the implementation within \texttt{scattering}\footnote{\url{https://github.com/SihaoCheng/scattering_transform}\href{https://github.com/SihaoCheng/scattering_transform}{}}   to calculate the SWT coefficients. We use the Morlet filter~\cite{10.5555/1525499}, which is local in both real and Fourier space. We calculate ST1 and ST2 with 6 spatial scales ($j_1$ and $j_2$) and 4 rotations. For the effects of different index and the relation to the images, see Ref.~\cite{2020MNRAS.499.5902C}.

\begin{figure*}
\centering
\includegraphics[width=1.99\columnwidth]{figs/ST1_tomo_v1_w_noise.pdf}
\includegraphics[width=1.99\columnwidth]{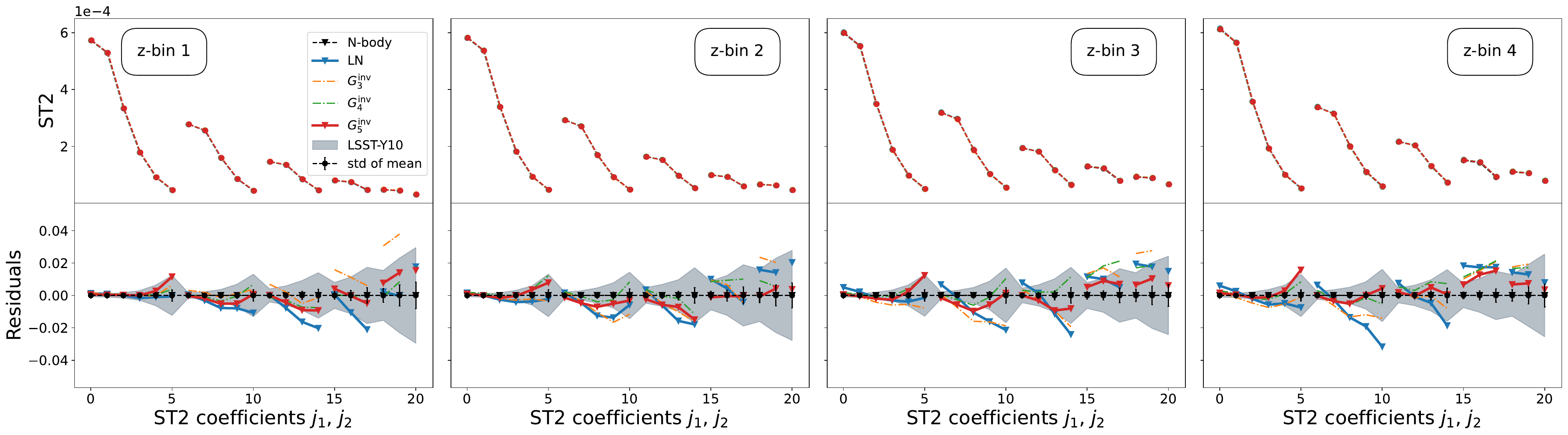}
\caption{The first and second Scattering Wavelet Transform coefficients ST1 and ST2. The index $j_1$ is a measure of scale. For ST2 in the lower panel, the scale increases for every few data points (same scale within connected lines). In the upper panels for ST1 and ST2, the different models are nearly indistinguishable by eye, but the residual panels show the failure of the simpler models relative to the LSST uncertainty bands. As expected, the large sale correlations are more accurate while small differences appear on the smaller scales.}
\label{fig:ST_tomo_comparison} 
\end{figure*}

As shown in Fig.~\ref{fig:ST_tomo_comparison}, the fitting formula that provides better PDF predictions also yields better SWT statistics. The fact that a few coefficients differ slightly suggests that SWT captures more non-Gaussian information than the point-transformation method can produce, such as filaments. Note that the LogNormal fields produce inaccurate SWT coefficients. For ST1, the 5-parameter maps gave results consistent with the LSST-Y10-like data, while the LogNormal maps gave biased results on certain coefficients. The fractional difference for 5-parameter maps is 4-6 times better than the LogNormal case. Note that in this case, the 5-parameter maps give better consistency than other parameterizations. For ST2, there are certain coefficients that the GPTG maps failed to reproduce. This result suggests that this kind of summary statistics could provide additional non-local information.

\subsection{Minkowski Functionals}\label{sec:Mink_Func}
\begin{figure*}
\centering
\includegraphics[width=1.99\columnwidth]{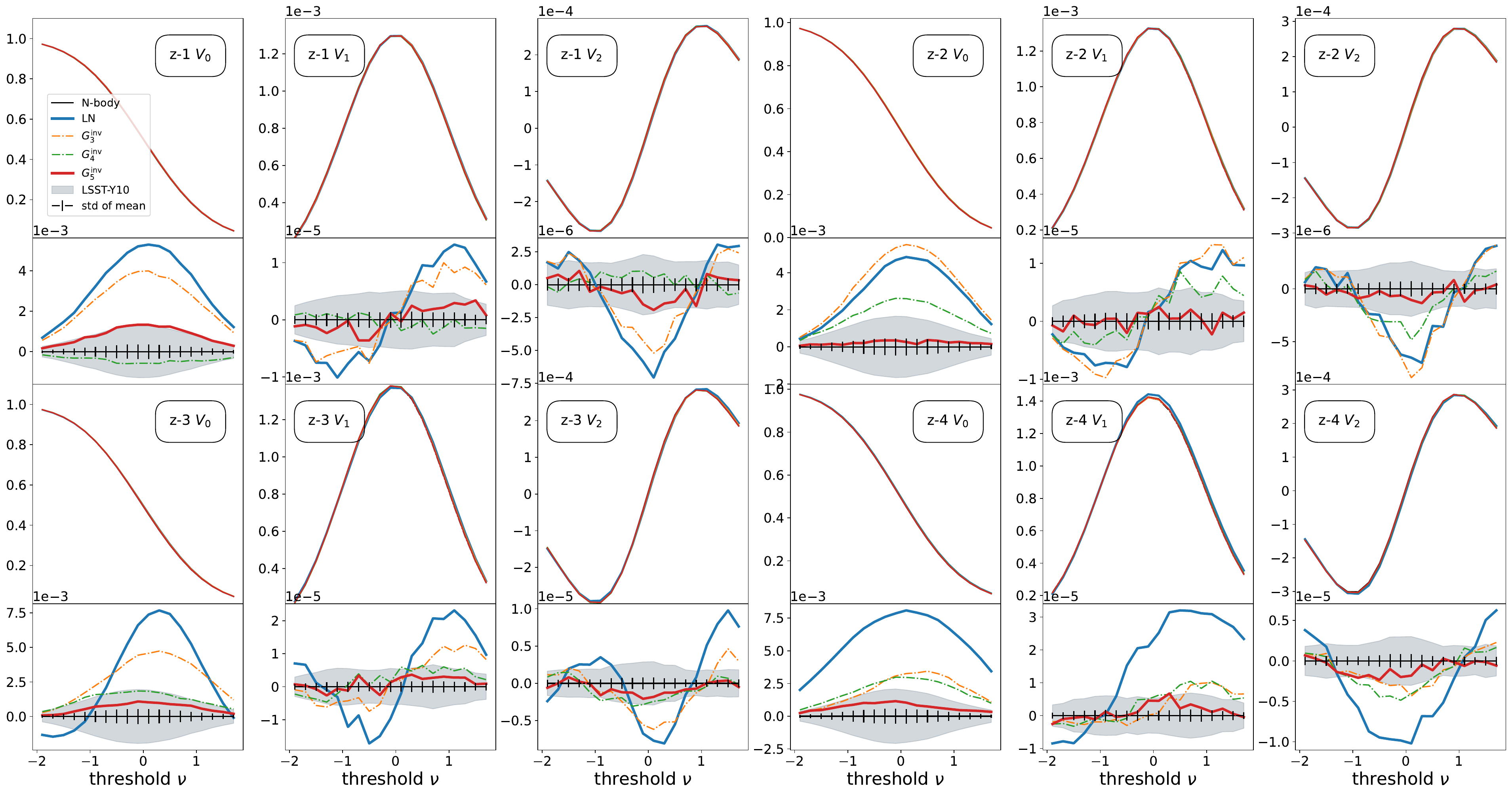}
\caption{Comparison of Minkowski Functionals (MF) for each redshift bin. For ease of visibility, we plot the difference $V_{\rm PT}- V_{\rm N-body}$ rather than the fractional difference used in other part of the paper. The MF are measured directly without any additional smoothing.} 
\label{fig:MF_comparison_tomo} 
\end{figure*}
\begin{figure*}
\centering
\includegraphics[width=1.99\columnwidth]{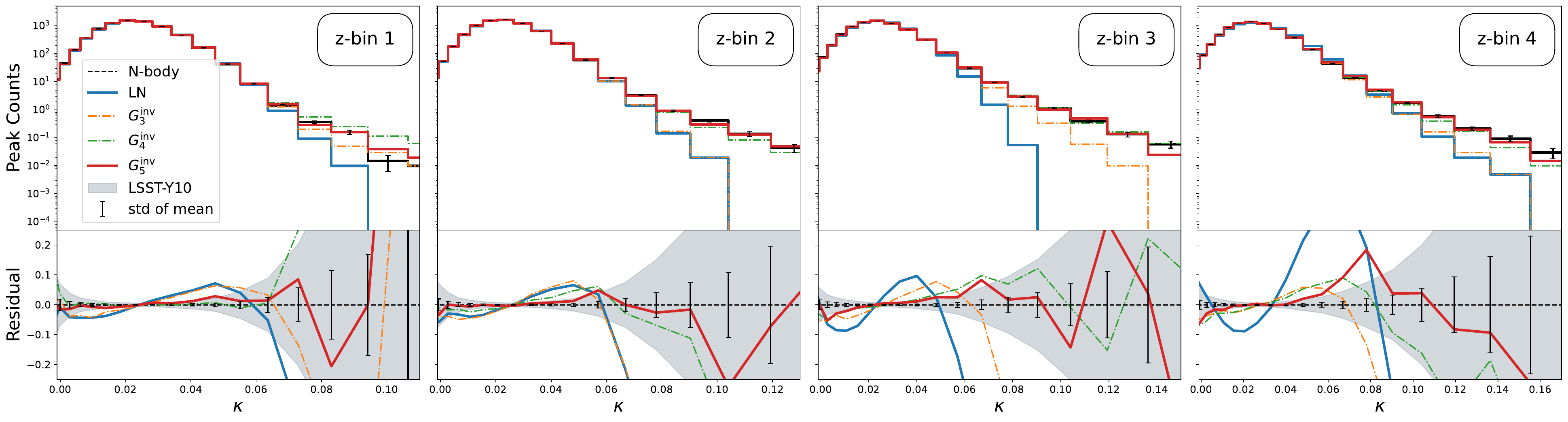}
\caption{Comparison of peak counts. Note that the number counts are shown on a log scale (on a linear scale the LogNormal map would appear to be the same, but the tails are very different). Peak counts are  very sensitive to small-scale modeling; we can only achieve the desired accuracy for our 5 parameter model with the power spectrum method introduced in Sec.~\ref{sec:power_spectrum_for_generation}. } 
\label{fig:peak_counts_comparison_tomo} 
\end{figure*}

The three Minkowski functions for $\kappa$ maps can be defined as follows:
\begin{equation}\label{eq:def_MF0}
\begin{aligned}
V_0(\nu)  &=\frac{1}{A} \int_{Q_v} \mathrm{~d} a \\
        &=\frac{1}{A} \int_A \mathrm{~d}^2 \boldsymbol{\vartheta} \mathcal{H}(\kappa-\nu \sigma) , \\   
\end{aligned}
\end{equation}

\begin{equation}\label{eq:def_MF1}
\begin{aligned}
V_1(\nu)  &=\frac{1}{4 A} \int_{\partial Q_\nu} \mathrm{~d} l \\
        &=\frac{1}{4 A} \int_A \mathrm{~d}^2 \boldsymbol{\vartheta} \delta_{\mathrm{D}}(\kappa-\nu \sigma) g(\kappa) , \\
\end{aligned}
\end{equation}

\begin{equation}\label{eq:def_MF2}
\begin{aligned}
V_2(\nu)  &=\frac{1}{2 \pi A} \int_{\partial Q_\nu} \mathcal{K} \mathrm{d} l \\
        &=\frac{1}{2 \pi A} \int_A \mathrm{~d}^2 \boldsymbol{\vartheta} \delta_{\mathrm{D}}(\kappa-\nu \sigma) h(\kappa) , 
\end{aligned}
\end{equation}
where $A$ is the area of the map and  $Q_\nu=\left\{\left(\vartheta_1, \vartheta_2\right) \mid \kappa\left(\vartheta_1, \vartheta_2\right) / \sigma \geq \nu\right\}$ is the excursion set, defined as pixels with values greater than the threshold $\nu$. $\mathcal{K}$ denotes the geodesic curvature of the excursion set boundary $\partial Q_\nu$. By definition (first equality), the three coefficients describe the area, boundary length, and genus (representation of connectivity) of the excursion set. The integral over the boundary $\partial Q_\nu$ can be carried to the surface integral with approximate finite derivatives $g(\kappa)$ and $h(\kappa)$ defined as:
\begin{equation}
\begin{aligned}
& g(\kappa)=\sqrt{\kappa_{,1}^2+\kappa_{, 2}^2} \\
& h(\kappa)=\left(\frac{2 \kappa_{, 1}^{\ } \kappa_{, 2}^{\ } \kappa_{, 12}^{\ }-\kappa_{, 1}^2 \kappa_{, 22}^{\ }-\kappa_{, 2}^2 \kappa_{, 11}^{\ }}{\kappa_{, 1}^2+\kappa_{, 2}^2}\right),
\end{aligned}
\end{equation}
In this work, we use the implementation in \texttt{LensTools}\footnote{\url{https://github.com/apetri/LensTools}\href{https://github.com/apetri/LensTools}{}} to calculate the Minkowski functionals. The results are shown in Fig.~\ref{fig:MF_comparison_tomo}. We see that the generalized point transformations give much better agreement than LogNormal. The $V_1(\nu)$ parameter shows the most discrepancy, which is expected because it captures structures in the image that are hard to produce by a local transformation of a Gaussian map. We did not show the results of a smoothed map, but they agree with the $N$-body maps better due to less non-Gaussinianity at larger scales. The 5-parameter model reproduces consistent results within the error of an LSST-Y10-like observation. Other parameterization still failed to generate accurate results although better than the LogNormal model, which in the worst case gives a $10 \sigma$ level bias as required by the LSST-Y10 observation.

\subsection{Peak Counts}\label{sec:peak_counts}

Peak counts are another powerful and efficient statistics for extracting non-Gaussian information~\cite{2000ApJ...530L...1J, 2009ApJ...698L..33M, 2010PhRvD..81d3519K, Liu:2016xjb, 2020ApJ...897...14C, Fluri:2018fpg, 2018MNRAS.474..712M, 2021MNRAS.506.1623H, 2022MNRAS.511.2075Z, 2023arXiv230810866M}. The number of local maxima from a map can arise from either a single massive halo or a superposition of multiple halos\cite{Liu:2016xjb}. The counting of peaks requires the application of a spatial filter (either explicitly or through pixelization), and multiple filters are usually used to get information at various scales. As a more stringent test to the method presented in this work, we compare the peak count statistics directly to the $\kappa$ maps without smoothing. We define peaks as the pixels that have greater values than all of its eight neighbors. Pixel-scale peaks are the most stringent tests because they require the inverse-Gaussianization functions to be accurate to the smallest scale. 

The results are shown in fig.~\ref{fig:peak_counts_comparison_tomo}. We plot the histograms of peaks in log scale and the fractional difference to visualize both the difference of the extreme tails and near the center. We notice that the 5-parameter model peaks are within the $1 \sigma$ range of the LSST-Y10 survey, whereas the LogNormal failed to generate peaks with high $\kappa$ values and biased even before the right tail.

\section{Applications and Limitations}\label{sec:applications_examples}

\subsection{Generalization  to Varying  Cosmology}\label{sec:vary_cosmology}
\begin{figure*}
\centering
\includegraphics[width=0.64\columnwidth]{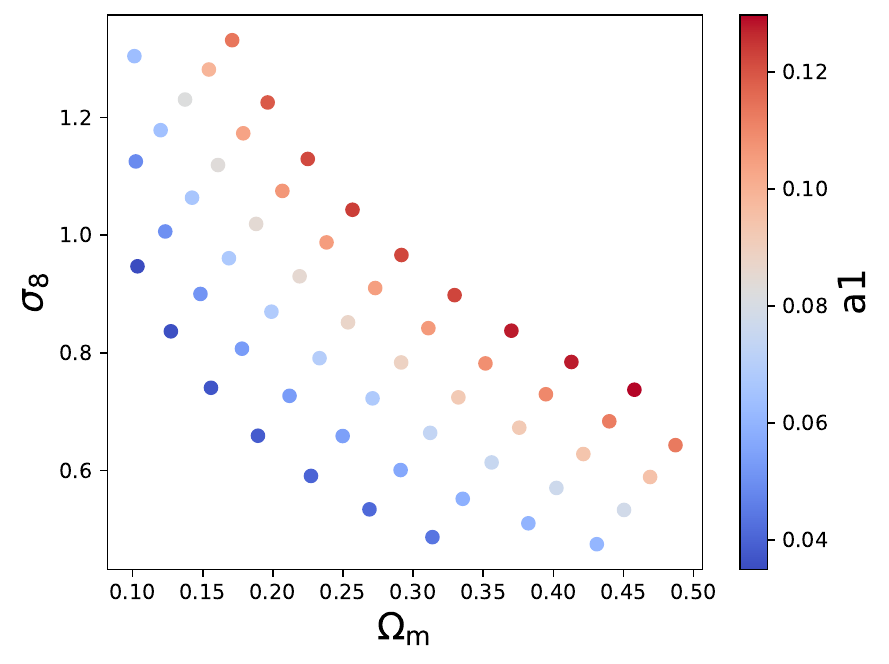}
\includegraphics[width=0.64\columnwidth]{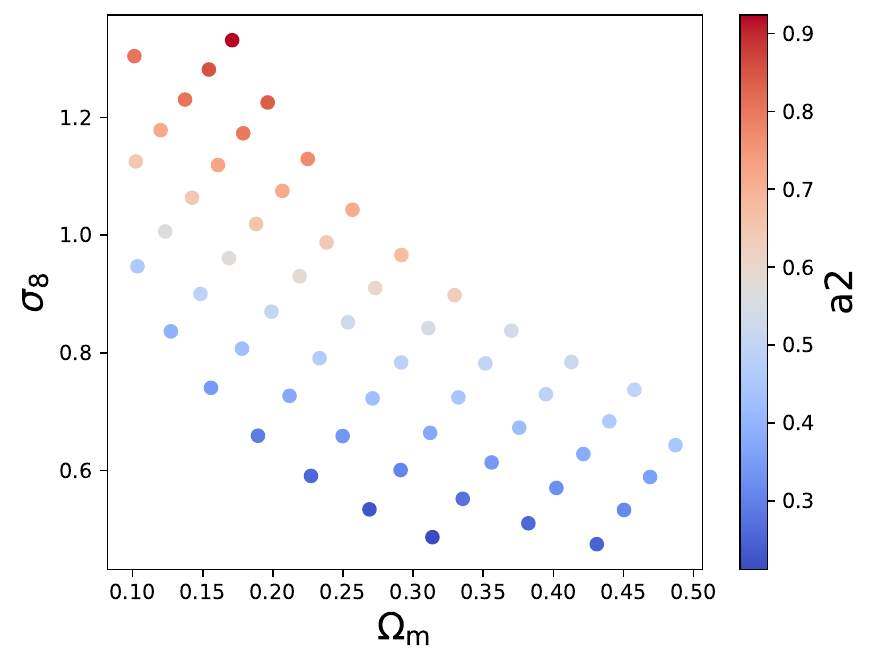}
\includegraphics[width=0.66\columnwidth]{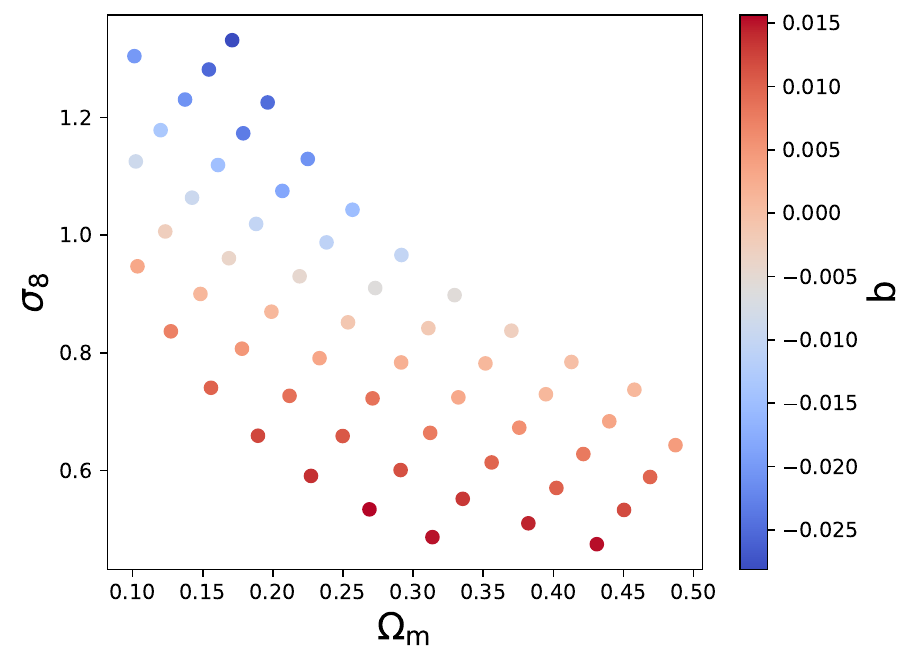}
\caption{The three panels show how $a_1$, $a_2$, and $b$ of the 5-parameter model vary with cosmological parameters. The different degeneracy directions of the three parameters illustrate how such an analytical model can be made interpretable. See Sec.~\ref{sec:vary_cosmology} for a detailed discussion.
}
\label{fig:cosmology_scatter}
\end{figure*}
One of the main advantages of describing the map generation process with only a few parameters is that it is more readily generalized across cosmologies. We show this by repeating the fitting procedure at different $\Omega_\mathrm{m}$ and $\sigma_8$ values and find that the 5-parameter function is flexible enough to describe different Gaussianization function. We find that the $a_1$ parameter in the 5-parameter model varies smoothly (except for small fluctuations at the upper boundary) and resembles the scaling of $S_8 \equiv \sigma_8 (\Omega_\mathrm{m}/0.3)^{1/2}$.  The $a_2$ parameter resembles the scaling of cluster number counts. This is interesting but surprising as  $a_2$ controls the large amplitude tail of the inverse Gaussianization function. This is an example of how such a generative model can be made more interpretable than say neural network-based models. 

We note that the $x_0$ and $t$ parameters are insensitive to the cosmological parameters at a given tomographic bin. We can thus redo the fitting procedure to reduce the degrees of freedom to 3. As an example, we show the third tomographic bin fitting in Fig.~\ref{fig:cosmology_scatter}. We can create a simple emulator from cosmology to the point transformation parameters by fitting a double power law~\footnote{Note that for numerical stability we fit the parameters to a rescaled convergence map $20 \kappa$.}:
\begin{equation}\label{eq:power_law_at_cosmo}
\begin{aligned}
    a_1 &= -0.0944 \sigma_8^{0.2443} \Omega_\mathrm{m}^{0.4860} \\
        & \quad + 0.4976 \sigma_8^{0.7385} \Omega_\mathrm{m}^{1.1450} \\
    a_2 &= 0.9459 \sigma_8^{0.7461} \Omega_\mathrm{m}^{1.7851} \\
        & \quad + 0.3193 \sigma_8^{-0.0378} \Omega_\mathrm{m}^{1.1450} \\
    b   &= -0.1180 \sigma_8^{0.4068} \Omega_\mathrm{m}^{1.0671} \\
        & \quad + 0.1081 \sigma_8^{0.1828} \Omega_\mathrm{m}^{0.0559} - 0.03\\
\end{aligned}
\end{equation}
We tested that such a simple emulator would give negligible errors in the center of the $\Omega_\mathrm{m} - \sigma_8$ space. At the boundary, it is slightly biased but unlikely to bias a realistic cosmological analysis. In any case the biases can be eliminated with a slightly more sophisticated emulator such as Gaussian Process Regression. Since the main purpose of this paper is not to test applications of cosmological inference, we defer detailed studies to future work.


\subsection{Estimating Covariance Matrices}\label{sec:covmat_estimation}
\begin{table*}[t]
\renewcommand{\arraystretch}{1.1}
\centering
\begin{tabular}{|c|c|c|c|c|c|}
\hline
Summary Statistics           & Degrees of Freedom & $\Delta \chi^2 (z_1)$     & $\Delta \chi^2 (z_2)$   & $\Delta \chi^2 (z_3)$  & $\Delta \chi^2 (z_4)$ \\ \hline



LogNormal PS      &   16               &   $0.18\pm3.11$   & $0.61\pm3.03$      &  $0.96\pm2.78$     & $2.30\pm3.91$  \\ \hline
5-parameter PS   &   16               &   $0.16\pm2.08$   & $1.44\pm2.56$      &  $2.08\pm2.72$     & $2.08\pm3.04$  \\ \hline
\multicolumn{6}{|l|}{\textbf{\textsc{}}}                \\ \hline
LogNormal ST2     &   21               &   $2.15\pm2.89$   & $3.53\pm3.57$       &  $7.55\pm6.64$     & $0.36\pm2.54$  \\ \hline
5-parameter ST2  &   21               &   $0.96\pm2.15$   & $1.91\pm2.49$        &  $0.87\pm2.31$     & $0.46\pm2.23$  \\ \hline

\multicolumn{6}{|l|}{\textbf{\textsc{}}}                \\ \hline
LogNormal MFV1     &   19               &   $1.23\pm2.26$   & $1.21\pm2.23$       &  $1.02\pm2.38$    & $0.66\pm2.16$  \\ \hline
5-parameter MFV1  &   19               &   $1.53\pm2.28$   & $0.61\pm1.98$        &  $0.64\pm1.84$   & $0.52\pm2.12$  \\ \hline

\end{tabular}
\caption{The expected $\chi^2$ shift of the covariance matrix estimated from LogNormal and 5-parameter models (mean $\pm$ standard deviation), compared to  $N$-body simulations at LSST-Y10 noise level. 
The deviation is significantly smaller for the best GPTG method compared to the LN for the higher-order statistics. An interesting case is that the power spectrum covariance is better estimated with the LN model -- likely due to the analytical LN formula being more accurate than the numerical method discussed in Sec.~\ref{sec:power_spectrum_for_generation}.}
\label{table:chi2_shift}
\end{table*}

LogNormal maps have previously been used in weak lensing and galaxy clustering analysis to estimate covariance matrix~\cite{Friedrich:2015nga, DES:2017qwj}. Although it is known they would not give accurate enough estimations for precision cosmology, it is still a useful tool to test the pipeline and other systematic effects, especially when a large number of maps are needed. A first approximation of covariance can also be used for data compression~\cite{Park:2024pxd}. In this section, we are interested to see if a more precise modeling could give a better estimation and thus mitigate possible bias.

For this section, we test the shape noise level corresponding to LSST-Y10 $n_\mathrm{eff}=27$. We utilize all 50 simulations at the fiducial cosmology of \texttt{DarkGridV1}. We project them down to 600 non-overlapping patches. Due to the limited number of N-body simulations, we test more GPTG maps to give consistent covariance matrix estimation (sub-percent in diagonal elements). We therefore expect 600 realizations to be sufficient in covariance estimation for data vectors of length $\sim 20$.

The covariance matrix $\mathbf{C}$ estimated is commonly used in Bayesian parameter inference, we are thus interested in the possible shift of $\chi^2$ using two difference covariance matrices. The data vector space $\chi^2$ for a data vector $\boldsymbol{D}$ can be written as $\chi^2=(\boldsymbol{D}-\boldsymbol{M})^{\mathrm{T}} \mathbf{C}^{-1}(\boldsymbol{D}-\boldsymbol{M})$ where $\boldsymbol{M}$ is the model data vector from measurement or a fiducial data vector used for synthetic test. Then if we use another "false" covariance matrix $\mathbf{C}_1$, when assuming the data vector differences follow a Gaussian distribution $(\boldsymbol{D}-\boldsymbol{M}) \sim \mathcal{N} (0, \mathbf{C}_0)$, the chi-square shift $\Delta\chi^2$ has mean and variance:
\begin{equation}
\begin{aligned}
&\begin{aligned}
E\left[\Delta \chi^2\right] & =\operatorname{Tr}\left(\mathbf{C}_1^{-1} \mathbf{C}_0\right) -N_D \\
\operatorname{Var}\left[\Delta \chi^2 \right] & =\left[2 N_D+2 \operatorname{Tr}\left(\mathbf{C}_1^{-1} \mathbf{C}_0 \mathbf{C}_1^{-1} \mathbf{C}_0\right)-4 \operatorname{Tr}\left(\mathbf{C}_1^{-1} \mathbf{C}_0\right)\right] 
\end{aligned}\\
\end{aligned}
\end{equation}
In our case we treat the covariance estimated from the $N$-body sims as "truth" $\mathbf{C}_0$ and that from LogNormal or 5-parameter model as "false" $\mathbf{C}_1$. 

The results are shown in Table.\ref{table:chi2_shift}. We choose the power spectrum of the patch (log binned to 16 data points), the second coefficient of scattering transform (ST2), and the second Minkowski functional MFV1 as examples of 2-point and beyond 2-point statistics. As shown in the table, the 5-parameter model gives on average a factor of 2 better estimations than the LogNormal maps, especially for non-Gaussian statistics like MFV1. A small difference of $\chi^2$ less than 3 also means that the two covariances are unlikely to give noticeably different results in parameter posteriors. Although using the maps directly does not meet the conventional criterion of $\Delta \chi^2 < 1$, the GPTG maps can be used as a better approximation than the LogNormal maps.

\subsection{Limitations of the Local Point Transformations}\label{sec:limits_of_point_transform}
\begin{figure*}
\centering
\includegraphics[width=2.00\columnwidth]{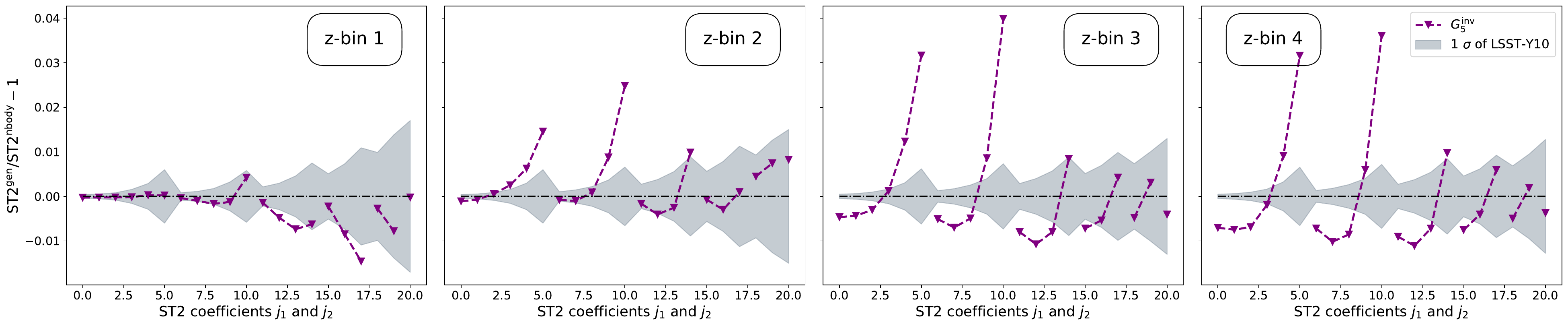}
\includegraphics[width=2.00\columnwidth]{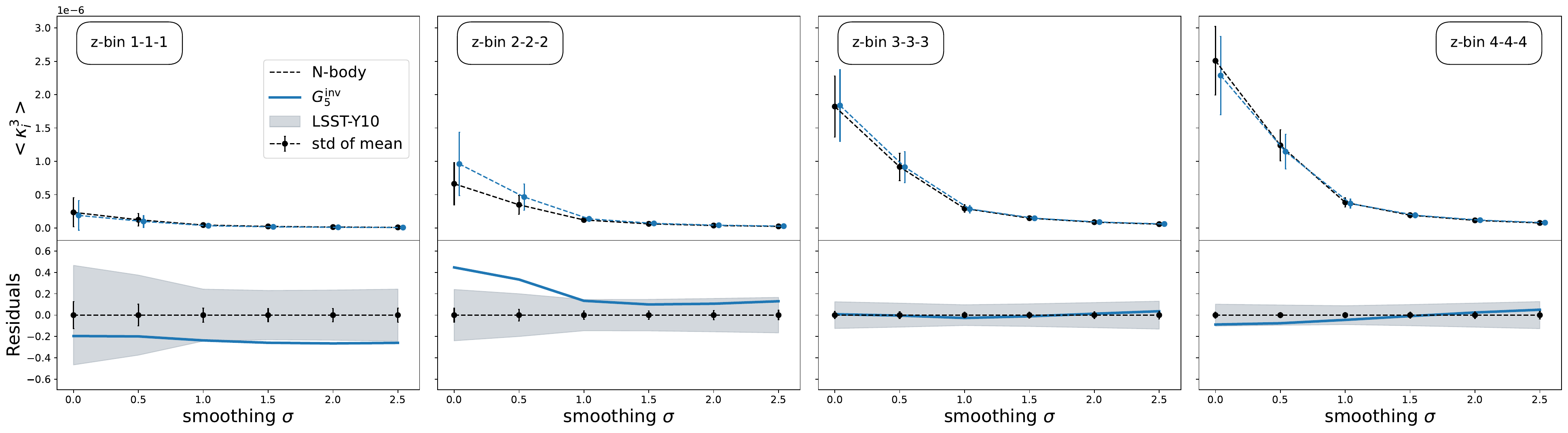}
\caption{Fractional difference of selected higher order statistics with a higher resolution map, 3.4 arcmin rather than the default 6.9 arcmin. At LSST-Y10 noise level, the generated higher resolution maps are more visually different from the $N$-body simulations. We show here two higher order statistics:  the ST2 statistics which is most sensitive to the difference as evident in the plot (note that the difference is still within five percent), and the third moments which are typical of other statistics that are still within the uncertainty band. 
However, at this smaller scale, we also need to consider baryonic feedback effects and thus we restrict ourselves to the 6.9 arcmin resolution maps in the rest of the paper.}
\label{fig:ST_2_T17sims} 
\end{figure*}
In this section, we present the accuracy of map generation at a higher angular resolution of 3.4 arcmin. We perform the test with the Takahashi17 simulations suite~\cite{Takahashi:2017hjr} as mentioned in Sec.\ref{sec:simulations}. We first redo the validations on summary statistics as in the~\texttt{DarkGridV1} using the low resolution (6.9 arcmin) and found it gives consistent results.

As an example, we show the scattering coefficients ST2 and third moments in Fig.~\ref{fig:ST_2_T17sims}. Compared to previous results at 6.9 arcmin resolution as in Fig.~\ref{fig:ST_tomo_comparison}, the ST2 results are significantly biased at LSST-Y10 noise level. However, since at this scale, we need to consider the effects of baryons, we refrain from producing maps with angular resolutions higher than 6.9 arcmin per pixel.

\section{Discussion and Conclusions}\label{sec:conclusions}

We have generalized the framework of LogNormal models to produce more accurate tomographic weak lensing convergence maps. By testing several general point-transformed Gaussian (GPTG) functions and calibrating them with $N$-body simulations, we have shown that modeling both the PDF and 2-point correlation function simultaneously can yield maps with accurate higher-order statistics. This finding is validated through various summary statistics: third-order moments, Scattering Wavelet Transform, Minkowski Functionals, and peak counts. Our analytical transform matches simulations within the statistical accuracy of the LSST-Y10 survey.  

Our method, particularly the more flexible 5-parameter model, offers advantages over black-box machine learning models due to its interpretability and ease of extension to new parameter spaces. Other applications include estimating covariance matrices and novel inference methods like  Field-Level-Inference~\cite{Boruah:2023fph, Zhou:2023ezg}, where its computational ease can make a critical difference. While we have not explored these potential applications, we anticipate that our method will serve as a useful analytical alternative to LogNormal models, especially when testing pipelines related to non-Gaussian statistics.

The potential applications of the GPTG in machine learning investigations are also intriguing. Previous studies have relied heavily on LogNormal maps due to their computational efficiency and ease of validation. For instance, Ref.~\cite{Akhmetzhanova:2023hiy} studied the self-supervised learning method using LogNormal maps. Ref.~\cite{Sharma:2024pth} explored the idea of transfer learning where they train/pre-train the neural network with LogNormal maps and test/fine-tune with $N$-body simulations. While they found no improvement in transfer learning, this could be attributed to the limited accuracy of LogNormal maps. Ref.~\cite{vonWietersheim-Kramsta:2024cks} used LogNormal maps to test a series of interesting systematics in simulation-based inference. A comparison between simulation-based inference and field-level inference~\cite{Lanzieri:2024mvn} in weak lensing is also an interesting application. All these examples, which involve understanding and applying machine learning methods in cosmology, could benefit from the more accurate analytical generative model presented in this work.

Despite the promising accuracy of our method in modeling convergence maps, we expect the accuracy to suffer when applied to higher-resolution maps with stronger nonlinear structures. The point-transformation method's effectiveness improves due to the smearing of these structures when integrating over the lensing kernel. For modeling three-dimensional density shells, we anticipate larger discrepancies. Another caveat of this work is the lack of intrinsic alignment modeling. The non-linear alignment (NLA) model~\cite{Bridle:2007ft} could in principle be added to the GPTG maps but more complicated models such as the tidal alignment and tidal torquing model (TATT)~\cite{PhysRevD.100.103506} would require modeling the density evolution directly. For baryon contamination to the $\kappa$ field, since we only applied the method to scales above 7 arcmin, the baryonic effect may not be the major systematics. Nevertheless, it would be interesting to see how the GPTG method changes when calibrating with baryonified maps or hydro-simulations. 

Potential solutions do exist for applications to higher resolutions, such as integrating a filter convolution before or after the point transform. CNN-based generative models have shown promise in this regard~\cite{Lanzieri:2022zvv, Lanzieri:2023ftk,Ding:2024owq}. Based on our findings, a small latent space of Neural Networks might suffice, potentially describable by exact analytical equations.

Looking beyond the convergence field, an analytical generative model can be found useful in 3D galaxy clustering~\cite{Ramanah:2018eed}. Galaxy clustering measurements generally have a higher signal-to-noise ratio but are harder to model because they are biased tracers of dark matter. A better density model, such as the method in this paper, could be a starting point. Together with modeling halos or galaxies, a generative model could help us model the physical process and statistical properties of the universe with better accuracy. While machine learning methods like diffusion models~\cite{Cuesta-Lazaro:2023zuk} have explored this avenue, our analytical approach offers a valuable alternative and validation tool.

In conclusion, we have presented a step forward in analytical and accurate modeling of lensing convergence maps. Our approach is useful as a generative model as well. By bridging the gap between simple LogNormal models and complex machine learning approaches, we provide a robust, interpretable, and generalizable method for cosmological studies. In future research, we will explore the applications of this method to various aspects of cosmological modeling and analysis.

\section*{Acknowledgements}

We thank Mike Jarvis and Shivam Pandey for useful discussions. We are especially grateful to Tomek Kacprzak and his collaborators for generating and making available the \textsc{DarkGrid} simulations used in this work. This research used resources of the National Energy Research Scientific Computing Center (NERSC), , operated under Contract No. DE-AC02-05CH11231. K.Z. and B.J. are partially supported by NASA funds for the Open Universe project and the Roman mission. G.B., S.B., and B.J. are partially supported by the US Department of Energy grant DE-SC0007901.

\bibliography{ref_short.bib}

\appendix

\section{Details of PDF fit implementations}\label{sec:pdf_fit_details}

Fitting the PDF of a convergence field can be numerically challenging. The absence of an analytical inverse function prevents us from deriving a PDF function of the general inverse-Gaussianization function. In this work, we employ a straightforward method: we start with a 1D Gaussian sample $x$ and determine the parameters that minimize the loss function between $N$-body simulations and $G^{-1}(x)$. While Kullback-Leibler (KL) divergence is a natural choice for the loss function in this context, we found no significant difference when using the root mean square error. When running for different cosmologies, we start with the same initial guess and iteratively run the \texttt{Powell}~\cite{10.1093/comjnl/7.2.155} minimizer until the loss meets the threshold. The minimizer, however, cannot explore the full parameter space, even with only five dimensions. Although an iterative approach could make the fitting more accurate for a single PDF, we maintain the more general minimizer settings to avoid bias when varying tomographic bins or cosmologies. A more sophisticated algorithm, such as Monte Carlo sampling, could be incorporated in a future study.

It's worth noting that depending on the choice of the loss function, the LogNormal function can be scaled to better fit the high $\kappa$ tails, but this would compromise the fit near the center because the exponential function lacks the freedom to fit both tails. Whether the LogNormal maps are good enough or not depends on the use case and noise level. We do not claim that the LogNormal field used in this paper is the optimal case, but we expect the main conclusion of this work to be unchanged, namely that by modeling the PDF  better, the generated point-transformed maps will have better higher-order statistics.

\end{document}